\documentclass[preprintnumbers,amsmath,amssymbm,prd]{revtex4}
\usepackage{epsfig}
\usepackage{graphicx}
\usepackage{amssymb}

\begin{document}
\title{Quasinormal resonances of rapidly-spinning Kerr black holes and the universal relaxation bound}
\author{Shahar Hod}
\affiliation{The Ruppin Academic Center, Emeq Hefer 40250, Israel}
\affiliation{ }
\affiliation{The Hadassah Institute, Jerusalem 91010, Israel}
\date{\today}

\begin{abstract}
\ \ \ The universal relaxation bound suggests that the relaxation times of 
perturbed thermodynamical systems is bounded from below by the simple 
time-times-temperature (TTT) quantum relation $\tau\times T\geq {{\hbar}\over{\pi}}$. 
It is known that some perturbation modes of near-extremal Kerr black holes in the regime $MT_{\text{BH}}/\hbar\ll m^{-2}$ 
are characterized by normalized relaxation times $\pi\tau\times T_{\text{BH}}/\hbar$ 
which, in the approach to the limit $MT_{\text{BH}}/\hbar\to0$, 
make infinitely many oscillations with a tiny constant amplitude around $1$ 
and therefore cannot be used directly to verify the validity of the TTT bound in the entire parameter space of the black-hole 
spacetime 
(Here $\{T_{\text{BH}},M\}$ are respectively the Bekenstein-Hawking temperature and the mass of the black hole, and $m$ is 
the azimuthal harmonic index of the linearized perturbation mode). 
In the present compact paper we explicitly prove that all rapidly-spinning Kerr black holes 
respect the TTT relaxation bound. In particular, using analytical techniques, it is proved 
that all black-hole perturbation modes 
in the complementary regime $m^{-1}\ll MT_{\text{BH}}/\hbar\ll1$ are characterized by relaxation times 
with the simple dimensionless property $\pi\tau\times T_{\text{BH}}/\hbar\geq1$.
\end{abstract}
\bigskip
\maketitle

%]

\section{Introduction}

Perturbed black-hole spacetimes are usually characterized by a relaxation phase which is 
dominated by exponentially decaying quasinormal oscillations whose discrete complex frequencies $\{\omega_n\}^{n=\infty}_{n=0}$
encode valuable information about the fundamental physical parameters (in particular, the mass and angular momentum) 
of the perturbed black hole \cite{Nollert1,Press,Cruz,Vish,Davis,QNMs}. 

The damped quasinormal oscillations of matter and radiation fields in perturbed black-hole spacetimes, 
which describe waves with outgoing boundary conditions at spatial infinity and ingoing boundary conditions 
at the absorbing black-hole horizon \cite{Detwe}, 
reflect the fact that most perturbations fields (fields that conform to existing no-hair theorems \cite{Whee,Car,Bek1,Chas,BekVec,Hart,Hod11}) are eventually scattered away to infinity or absorbed by the 
central black hole \cite{Notetails,Tails,Notescc,Hodrc,HerR}.

The timescale $\tau_{\text{relax}}$ associated with the relaxation dynamics of 
a newly born black hole may be determined by the imaginary part of the fundamental (least
damped) quasinormal resonant frequency:
\begin{equation}\label{Eq1}
\tau_{\text{relax}}\equiv 1/\Im\omega_0\  .
\end{equation}
Taking cognizance of the fact that black holes have thermodynamic and quantum properties \cite{Bektem,Hawtem}, 
a physically important question naturally arises: How short can the
relaxation time (\ref{Eq1}) of a newly born black hole be?

A mathematically compact answer to this physically intriguing question, 
which is based on standard ideas from classical thermodynamics and quantum information theory, 
has been suggested in \cite{HodTTT,Notetl}: 
\begin{equation}\label{Eq2}
\tau_{\text{relax}}\times T\geq{{\hbar}\over{\pi k_{\text{B}}}}\  ,
\end{equation}
where $T$ is the characteristic temperature of the thermodynamic physical system.

Intriguingly, remembering that the semi-classical Bekenstein-Hawking 
temperature \cite{Bektem,Hawtem,Noteunit} 
\begin{equation}\label{Eq3}
T_{\text{BH}}={{\kappa}\over{2\pi k_{\text{B}}}}\cdot\hbar\
\end{equation}
of black holes depends linearly on the quantum Planck constant (here $\kappa$ is the characteristic 
surface gravity of the black-hole horizon), 
one realizes that the time-times-temperature (TTT) quantum bound (\ref{Eq2}) provides a classical lower 
bound \cite{Notehbarcal} on the characteristic relaxation time of a newly born black hole. 
In particular, taking cognizance of Eqs. (\ref{Eq1}), (\ref{Eq2}), and (\ref{Eq3}), one realizes that 
all dynamically formed black holes should be characterized by (at least) one relaxation mode 
with the simple classical property  
\begin{equation}\label{Eq4}
\Im\omega_0\leq{{\pi T_{\text{BH}}}\over{\hbar}}={1\over 2}\kappa\  .
\end{equation}

\section{Relaxation dynamics of rapidly spinning Kerr black holes}
 
Interestingly, the upper bound (\ref{Eq4}) implies 
that newly born near-extremal black holes (with the property $T_{\text{BH}}\to 0$) 
should be characterized by extremely long relaxation times \cite{HodTTT,Hodan,Als}.
In particular, using the physically important and mathematically elegant 
Detweiler equation \cite{Detw}, 
which characterizes the complex resonant spectra of rapidly-spinning (near-extremal) Kerr black holes, 
one obtains the simple functional relation \cite{Hoddet,Gruzdet,Notehbar} 
\begin{equation}\label{Eq5}
\Im\omega_0=\pi T_{\text{BH}}\cdot\big\{1+C\cdot\sin[2\delta\ln(MT_{\text{BH}})]\big\}
\cdot[1+O(MT_{\text{BH}})]\
\end{equation}
for the co-rotating perturbation modes of near-extremal black holes, where $C(l,m)$ is a constant \cite{NoteCC}, $M$ is the 
black-hole mass, and the physical parameter $\delta(l,m)\in\mathbb{R}$ is closely related to the characteristic 
eigenvalue of the angular Teukolsky equation \cite{Teuk,Notedelta}. 

Before proceeding, it is important to stress the fact that the validity of the 
Detweiler resonance equation \cite{Detw}, and 
thus also the validity of the analytically derived expression (\ref{Eq5}) \cite{Hoddet,Gruzdet}, are restricted to the dimensionless 
regime \cite{Teuk,Starrv,Hodmmm}
\begin{equation}\label{Eq6}
m^2\cdot MT_{\text{BH}}\ll1\  ,
\end{equation}
where $m$ is the azimuthal harmonic index of the black-hole perturbation mode. 

From Eq. (\ref{Eq5}) one immediately finds that perturbation modes of rapidly-spinning 
Kerr black holes in the regime (\ref{Eq6}) are characterized by normalized relaxation times $\pi\tau\times T_{\text{BH}}$ 
that oscillate infinitely many times around $1$ as the extremal limit $MT_{\text{BH}}\to0$ is approached. 
One therefore deduces that, in the near-extremal $MT_{\text{BH}}\ll1$ regime, 
there are finite intervals of the black-hole temperature for which a perturbation mode in the dimensionless regime (\ref{Eq6}) 
cannot be used to prove the general validity of the TTT relaxation bound (\ref{Eq2}) [or equivalently, the upper 
bound (\ref{Eq4})] in black-hole physics \cite{Notenng}. 

The main goal of the present compact paper is to prove the physically important fact 
that {\it all} rapidly-spinning Kerr black holes respect the TTT relaxation bound (\ref{Eq2}). 
To this end, we shall study below the linearized relaxation dynamics of newly born near-extremal Kerr black holes. 
In particular, using analytical techniques, we shall explicitly prove that composed black-hole-field perturbation modes 
in the eikonal large-$m$ regime
\begin{equation}\label{Eq7}
m^{-1}\ll MT_{\text{BH}}\ll1\
\end{equation}
are characterized by relaxation times that respect the TTT relaxation bound. 

\section{Resonant relaxation spectra of rapidly spinning Kerr black holes in the eikonal large-frequency regime}

Using a physically intuitive and mathematically elegant analysis, Mashhoon \cite{Mash} has presented a simple
analytical technique for calculating the quasinormal resonance spectra of black holes in the large-frequency 
(geometric-optics) regime. In particular, the physical idea presented in \cite{Mash,Goeb} 
is to relate, in the eikonal large-frequency regime, the real part of the black-hole quasinormal frequencies 
to the angular velocity which characterizes the motion of massless particles along the 
equatorial null circular geodesic of the black-hole spacetime and to relate the imaginary 
part of the black-hole quasinormal frequencies to the instability timescale 
which characterizes the gradual leakage of the perturbed massless particles (null rays) from the unstable 
null circular geodesic of the black-hole spacetime. 

In particular, using a perturbation scheme for the instability which characterizes the dynamics 
of equatorial null circular geodesics in the Kerr black-hole spacetime, 
Mashhoon has provided the remarkably compact formula \cite{Mash}
\begin{equation}\label{Eq8}
\omega_n =m\omega_+  -i(n+{1\over 2})\beta\omega_+\ \ \ \ ;\ \ \ \ n=0,1,2,...\ ,
\end{equation}
for the discrete quasinormal resonant spectra of spinning Kerr black holes. As emphasized in \cite{Mash}, 
this resonance formula is valid for perturbation modes in the eikonal large-frequency regime
\begin{equation}\label{Eq9}
l=m\gg1\  ,
\end{equation}
where $\{l,m\}$ are the (spheroidal and azimuthal) angular harmonic indexes which characterize the linearized field mode. 

The various terms in the resonance formula (\ref{Eq8}) have the following physical interpretations \cite{Mash}:
\newline
(1) The term 
\begin{equation}\label{Eq10}
r_{\text{ph}}(a/M)\equiv 2M\cdot\Big\{1+\cos\Big[{2 \over 3}\arccos\Big(-{{a}\over{M}}\Big)\Big]\Big\}\
\end{equation}
is the radius of the co-rotating equatorial null circular geodesic, where $\{M.a\}$ are 
respectively the mass and angular momentum per unit mass \cite{Noteaaa} of the central spinning Kerr black hole.
\newline
(2) The term
\begin{equation}\label{Eq11}
\omega_+(a/M)\equiv {{M^{1/2} \over{r^{3/2}_{\text{\text{ph}}}+aM^{1/2}}}}\
\end{equation}
is the Kepler frequency which characterizes the co-rotating equatorial null circular geodesic 
of the black-hole spacetime.
\newline 
(3) The dimensionless functional expression of $\beta=\beta(M,a)$ is given by \cite{Mash}
\begin{equation}\label{Eq12}
\beta(a/M)= {{(12M)^{1/2}(r_{\text{ph}}-r_+)(r_{\text{ph}}-r_-)}\over{r^{3/2}_{\text{ph}}(r_{\text{ph}}-M)}}\  ,
\end{equation}
where 
\begin{equation}\label{Eq13}
r_{\pm}=M\pm(M^2-a^2)^{1/2}\
\end{equation}
are the (outer and inner) horizon radii of the spinning Kerr black hole.

We shall henceforth focus our attention on rapidly-spinning (near-extremal) Kerr black holes, which are 
characterized by the simple dimensionless relation
\begin{equation}\label{Eq14}
{{a}\over{M}}=1-\epsilon\ \ \ \text{with}\ \ \ 0\leq\epsilon\ll1\  .
\end{equation}
From Eqs. (\ref{Eq13}) and (\ref{Eq14}) one finds the relation
\begin{equation}\label{Eq15}
{{r_{\pm}}\over{M}}=1\pm\sqrt{2}\sqrt{\epsilon}\mp{{\epsilon^{3/2}}\over{2\sqrt{2}}}+O(\epsilon^{5/2})\  .
\end{equation}

Substituting (\ref{Eq14}) into Eq. (\ref{Eq10}), one obtains 
the near-extremal (small-$\epsilon$) dimensionless functional expression
\begin{equation}\label{Eq16}
{{r_{\text{ph}}}\over{M}}=1+{{2\sqrt{2}}\over{\sqrt{3}}}\cdot\sqrt{\epsilon}+{4\over9}\cdot\epsilon+O(\epsilon^{3/2})\
\end{equation}
for the radius of the co-rotating equatorial null circular geodesic. 
Substituting Eqs. (\ref{Eq14}), (\ref{Eq15}), and (\ref{Eq16}) into Eqs. (\ref{Eq11}) and (\ref{Eq12}), one finds
\begin{equation}\label{Eq17}
M\omega_+={1\over2}-{{\sqrt{3}}\over{2\sqrt{2}}}\cdot\sqrt{\epsilon}+{7\over12}\cdot\epsilon+O(\epsilon^{3/2})\
\end{equation}
and
\begin{equation}\label{Eq18}
\beta=\sqrt{2}\cdot\sqrt{\epsilon}-{{4}\over{3\sqrt{3}}}\cdot\epsilon+O(\epsilon^{3/2})\  .
\end{equation}

Substituting Eqs. (\ref{Eq17}) and (\ref{Eq18}) into the resonance formula (\ref{Eq8}),
one obtains the dimensionless functional expression 
\begin{equation}\label{Eq19}
M\omega_n =m\cdot\Big({1\over2}-{{\sqrt{3}}\over{2\sqrt{2}}}\cdot\sqrt{\epsilon}+{7\over12}\cdot\epsilon\Big)-
i\big(n+{1\over2}\big)\cdot\Big({{\sqrt{\epsilon}}\over{\sqrt{2}}}-{{13}\over{6\sqrt{3}}}\cdot\epsilon\Big)+O(\epsilon^{3/2})\  .
\end{equation}
for the complex quasinormal resonant frequencies of rapidly-spinning (near-extremal) Kerr black holes in the eikonal 
large-frequency regime $l=m>>1$ [see Eq. (\ref{Eq9})].

It is physically interesting and mathematically convenient to express the analytically derived 
Kerr resonance spectrum (\ref{Eq19}) 
in terms of the black-hole Bekenstein-Hawking temperature \cite{Bektem,Hawtem} 
\begin{equation}\label{Eq20}
T_{\text{BH}}={{r_+-r_-}\over{4\pi(r^2_++a^2)}}=
{{1}\over{\pi M}}\cdot\Big[{{\sqrt{\epsilon}}\over{2\sqrt{2}}}-{{\epsilon}\over{2}}+O(\epsilon^{3/2})\Big]\
\end{equation}
and the black-hole angular velocity
\begin{equation}\label{Eq21}
\Omega_{\text{H}}={a \over {r^2_+ +a^2}}={{1}\over{M}}\cdot\Big[{1\over 2}-{{\sqrt{\epsilon}}\over{\sqrt{2}}}+
{{\epsilon}\over{2}}+O(\epsilon^{3/2})\Big]\  .
\end{equation}
Substituting (\ref{Eq20}) and (\ref{Eq21}) into Eq. (\ref{Eq19}), one finds the black-hole resonance formula 
\begin{equation}\label{Eq22}
\omega_n =\Big\{m\Omega_{\text{H}}\cdot\big[1+\pi(4-2\sqrt{3})MT_{\text{BH}}\big]-
i\big(n+{1\over2}\big)\cdot 2\pi T_{\text{BH}}\cdot\big[1-{{2\pi}\over{9}}(13\sqrt{3}-18)MT_{\text{BH}}\big]\Big\}
\cdot[1+O(MT_{\text{BH}})]\  .
\end{equation}

\section{Summary} 

Motivated by the suggested time-times-temperature (TTT) universal relaxation bound (\ref{Eq2}), we have studied, 
using analytical techniques, the quasinormal resonance spectra of near-extremal (rapidly-spinning) 
Kerr black holes. 

It was first noted that perturbation modes of near-extremal  
Kerr black holes in the regime $m^2\cdot MT_{\text{BH}}\ll1$ 
are characterized by normalized relaxation times $\pi\tau\times T_{\text{BH}}$ 
which, in the approach to the extremal limit $MT_{\text{BH}}\to0$, oscillate infinitely many times around $1$ 
and therefore cannot be used to verify the general 
validity of the suggested TTT relaxation bound (\ref{Eq2}) [or equivalently, the upper 
bound (\ref{Eq4})] in black-hole physics. 

We have therefore analyzed the black-hole relaxation spectra in the 
complementary regime $m^{-1}\ll MT_{\text{BH}}\ll1$. 
In particular, we have explicitly proved that the equatorial black-hole-field perturbation modes 
in the eikonal large-frequency regime (\ref{Eq7}) are characterized by the simple near-extremal relation 
%\cite{Noten0}
\begin{equation}\label{Eq23}
\Im\omega_0=
\pi T_{\text{BH}}\cdot\big[1-{{2\pi}\over{9}}(13\sqrt{3}-18)MT_{\text{BH}}\big]<\pi T_{\text{BH}}\ \ \ \ \text{for}\ \ \ \ 
m^{-1}\ll MT_{\text{BH}}\ll1\  .
\end{equation}
[It is worth noting that the functional relation (\ref{Eq23}) characterizes the fundamental ($n=0$) 
resonant mode of the analytically derived black-hole relaxation spectrum (\ref{Eq22})]. 
We therefore conclude that the eikonal relaxation modes (\ref{Eq23}), which characterize 
the composed black-hole-field system in the dimensionless large-frequency regime $m^{-1}\ll MT_{\text{BH}}\ll1$, guarantee that 
newly born near-extremal Kerr black holes respect the suggested TTT universal relaxation bound (\ref{Eq2}) \cite{Notenln}. 

\bigskip
\noindent
{\bf ACKNOWLEDGMENTS}
\bigskip

This research is supported by the Carmel Science Foundation. I would
like to thank Yael Oren, Arbel M. Ongo, Ayelet B. Lata, and Alona B.
Tea for helpful discussions.

%\newpage


\begin{thebibliography}{99}
\bibitem{Notescc} It is interesting to note that some non-decaying matter fields can 
be supported in asymptotically flat black-hole spacetimes without being 
absorbed into the central supporting black holes \cite{Hodrc,HerR}.
\bibitem{Notetails} It is worth mentioning that the late-time relaxation dynamics of perturbed black holes in 
asymptotically flat spacetimes is also characterized by inverse power-law decaying tails which 
reflect the late-time backscattering of dynamical perturbation fields by the small 
curvature potential at asymptotically far regions \cite{Tails}.
\bibitem{Notetl} It is interesting to note that the quantum relaxation bound (\ref{Eq2}) 
provides a quantitative formulation for the fundamental third law of thermodynamics.
\bibitem{Noteunit} We use natural units in which $G=c=k_{\text{B}}=1$.
\bibitem{Notehbar} For brevity, we shall henceforth use units in which $\hbar=1$.
\bibitem{NoteCC} The amplitude $C=C(l,m)$ depends on the angular harmonic indexes of 
the composed black-hole-field perturbation mode \cite{Gruzdet,Hoddet}. 
It is worth noting that one finds $|C|\ll1$ for most perturbation modes \cite{Gruzdet,Hoddet}.
\bibitem{Notedelta} It is worth noting that the simple relation (\ref{Eq5}) characterizes 
all near-extremal perturbation modes of rapidly-spinning Kerr black holes 
with the property $\delta^2>0$ \cite{Gruzdet,Hoddet}. 
For near-extremal perturbation modes 
with the property $\delta^2<0$ one finds the relation
$\Im\omega_0=\pi T_{\text{BH}}\cdot(1+2|\delta|)\cdot[1+O(MT_{\text{BH}})]>\pi T_{\text{BH}}$ \cite{Gruzdet,Hoddet}.
\bibitem{Notenng} It is interesting to note that numerical studies indicate that, for a given black-hole temperature, 
there is at least one black-hole-field perturbation mode in the regime (\ref{Eq6}) that conforms 
to the TTT relaxation bound (\ref{Eq4}) \cite{HodTTT,Gruzdet,Hoddet}. 
However, to the best of our knowledge there is no general argument which guarantees that all black holes must be 
characterized by this property and it is therefore physically important to present a more generic mathematical proof 
for the validity of the suggested 
TTT relaxation bound (\ref{Eq2}) [or equivalently, the upper bound (\ref{Eq4})] in black-hole physics.
\bibitem{Noteaaa} We shall assume $a\geq0$ without loss of generality.
\bibitem{Notenln} It is worth noting again that one learns from the functional relation (\ref{Eq5}) that some linearized
perturbation modes in the dimensionless regime (\ref{Eq6}) may not conform to the TTT relaxation bound (\ref{Eq2}) 
for particular black-hole temperatures. 
%It is important to emphasize that, due to the non-linear character of 
%the composed Einstein-matter field equations, the infinitely many black-hole-field excitation modes are 
%expected to be non-linearly coupled. Thus, 
The TTT relaxation bound asserts that a dynamically formed black hole should 
be characterized by (at least) one relaxation mode with the property (\ref{Eq4}). In the present compact paper we have explicitly 
proved that the resonant perturbation modes (\ref{Eq23}) of near-extremal Kerr black holes in the dimensionless regime (\ref{Eq7}) 
conform to the suggested TTT relaxation bound (\ref{Eq2}) [or equivalently, to the upper bound (\ref{Eq4})].
\end{thebibliography}
\end{document}